\documentstyle[12pt,aasms4]{article}

\newcommand{\oiii}{\hbox{[{\ion{O}{3}}]}}

\received{}
\accepted{}

\slugcomment{Submitted to the Astronomical Journal}

\lefthead{Wilson et al.}
\righthead{Hubble Space Telescope Imaging of the Circinus Galaxy}

\begin{document}

\title{Hubble Space Telescope Imaging of the Circinus
Galaxy\footnote{Based on observations with the NASA/ESA {\it Hubble
Space Telescope} obtained at the Space Telescope Science Institute, which
is operated by the Association of Universities for Research in Astronomy,
Inc., under NASA contract NAS 5-26555}}

\author{A. S. Wilson}
\affil{Space Telescope Science Institute, 3700 San Martin Drive,
Baltimore, MD 21218; awilson@stsci.edu}
\affil{and}
\affil{Astronomy Department, University of Maryland, College Park,
MD 20742; wilson@astro.umd.edu}

\author{P. L. Shopbell\altaffilmark{2}}
\affil{Astronomy Department, University of Maryland, College Park,
MD 20742; pls@astro.umd.edu}

\author{Chris Simpson}
\affil{Subaru Telescope, National Astronomical Observatory of Japan,
650 North A'ohuku Place, Hilo, HI 96720; chris@naoj.org}

\author{T. Storchi-Bergmann, F. K. B. Barbosa}
\affil{Instituto di Fisica, Universidad Federal do Rio Grande do Sul, 
Porto Alegre, RS 91501-970, Brazil; thaisa@if.ufrgs.br}

\and

\author{M. J. Ward}
\affil{Department of Physics \& Astronomy, University of Leicester,
University Road, Leicester, LE1 7RH, United Kingdom; mjw@star.le.ac.uk}


\altaffiltext{2}{Present address: Department of Astronomy, Mail Code 105-24,
California Institute of Technology, Pasadena, CA 91125; pls@phobos.caltech.edu}


\begin{abstract}
We present an HST imaging study of the nearby (4 Mpc distant)
Circinus galaxy, which contains the
nearest type 2 Seyfert nucleus and prominent circumnuclear
star formation. Images have been obtained in the [OIII]$\lambda$5007,
H$\alpha$
and H$_2$\,v=1--0\,S(1) emission lines, and in the
green (5470 \AA), red (8140 \AA) and near infrared (2.04 and 2.15
$\mu$m) continua. An image in the 
[Fe II] $\lambda$1.644 $\mu$m line has been taken with a
ground-based telescope. The [OIII] and
H$\alpha$ images reveal the detailed structure of the complex of streamers and
knots of
high-excitation gas, which extends out of the galaxy disk.
The morphology some 250 pc from the nucleus strongly suggests that the
high-excitation gas is concentrated on the surface of a hollow cone
with apex close to the nucleus. Such a structure may result through entrainment
of dense gas from a circumnuclear torus in the galaxy disk by a low
density, outflowing wind or jet. Within 40 pc of the nucleus, the
high-excitation gas takes the form of a striking, filled V-shaped structure
extending in the same direction as the larger scale high-excitation emission.
This `V' can be described as an ionization cone, though a matter-bounded
structure is also possible. The implied collimation of the ionizing photons
or gaseous outflow must occur within 2 pc of the apex of the cone, presumed
to be the location of the nucleus.

The H$\alpha$ image shows a complex structure of HII regions, including
the well known starburst ring of radius 150 - 270 pc. In addition, there is
a more compact (40 pc radius), elliptical ring of HII regions around the
`ionization cone'. We argue that this latter ring,
which we call the nuclear ring, is intrinsically 
circular and located in the plane of the galaxy disk. Much of the [Fe II]
emission is associated with this nuclear, star-forming ring and is presumably
powered by supernova remnants. Hot molecular hydrogen extends to within
10 pc of the nucleus, and possibly closer. The intrinsic infrared - optical
continuum colors in the inner regions of the Circinus galaxy
are, in many locations,
bluer than is typical of bulges,
indicating a relatively young stellar population is present.
We confirm the presence of a compact ($<$ 2 pc), very red nuclear source
in the K band. Its properties are consistent with a type 1 Seyfert nucleus
viewed through an obscuration of $A_V = 28 \pm 7$ mag.

\end{abstract}




%

\newpage
\section{Introduction}

The Circinus galaxy is a large, highly inclined ($i = 65^{\circ}$), spiral
galaxy of uncertain morphological
type
seen through a Galactic window with low extinction (A$_{V}$ = 1.5 $\pm$ 0.15
mag) at b = --4$^{\circ}$ (Freeman et al. 1977). It contains
both a type 2 Seyfert nucleus and a circumnuclear starburst and, in view of its
proximity (distance 4 $\pm$ 1 Mpc,
thus 1$^{\prime\prime}$ = 19 pc), has been intensively studied in recent
years. The Seyfert activity was first found through the discovery
of water vapor megamaser emission (Gardner \& Whiteoak 1982), a phenomenon
apparently associated exclusively with Seyfert or LINER-type active nuclei
(Braatz, Wilson \& Henkel 1997).
Individual water vapor maser spikes vary
on timescales as short as a few minutes (Greenhill et al. 1997) and
the masing gas
traces a thin accretion disk about 0.8 pc in radius with, in addition,
a significant population of masers that lie away from the disk and may be
in outflow (Greenhill et al. 1998). The X-ray spectrum below 10 keV
exhibits a flat continuum and a very prominent iron line, indicative of
Compton scattering and fluorescent emission from gas illuminated by
an obscured X-ray continuum source (Matt et al. 1996). Recent observations above
10 keV have revealed excess emission above an extrapolation of the 
0.1 - 10 keV spectrum suggesting that, at the higher energies,
the nucleus is seen directly through
a column density of $\sim$ 4 $\times$ 10$^{24}$ cm$^{-2}$ (Matt et al. 1999). 
Optical and near infrared spectrophotometry of the nucleus show a
typical Seyfert spectrum, including strong coronal lines (Oliva et al. 1994;
Maiolino et al. 1998).
Emission lines from highly ionized species are also found in a 2.5 - 45 $\mu$m
spectrum obtained with ISO
(Moorwood et al. 1996).
The gas is generally believed to be photoionized though
quite different photoionization
models (Moorwood et al. 1996; Binette et al. 1997; Oliva,
Marconi \& Moorwood 1999) can reproduce the observed emission-line ratios.
Optical spectropolarimetry reveals polarized and relatively broad (FWHM
$\sim$ 3,300 km s$^{-1}$) H$\alpha$ emission from a region $<$ 60 pc in
extent centered on the nucleus, indicating the presence of a `hidden' broad
line region (Oliva et al. 1998).

Highly ionized gas extends along the minor
axis of the galaxy, with a morphology that widens with increasing
distance from the nucleus and is reminiscent of the so-called `ionization cones'
seen in some Seyfert galaxies (Marconi et al. 1994). The `cone' in Circinus
does not, however, show the filled structure with
sharp, straight edges seen in some other
Seyferts (e.g. NGC 5728, Wilson et al. 1993) and which
strongly suggests
the gas is ionized by a shadowed nuclear uv source.
A complex of ionized filaments is found within the `cone'
and extending radially
from the nucleus out to distances of 1 kpc, as well as arcs suggestive
of bow shocks at the termini of some of these filaments
(Veilleux \& Bland-Hawthorn 1997; Elmouttie et al. 1998a). The filamentary
gas appears to be flowing outwards from the nucleus at velocities of 
200 km s$^{-1}$ or higher.
In the radio continuum (Elmouttie et al. 1995, 1998b), there is
a compact ($<$ 20 pc diameter) flat spectrum nuclear source plus
an extended cross-like structure. One arm of the cross extends along the
galaxy disk, and is believed related to star formation activity, while the
other extends orthogonal to it. The orthogonal radio features are strongly
polarized and
comprise a bi-symmetric plume and edge-brightened
lobes. 

A circumnuclear star-forming ring of diameter $\simeq$ 200 pc is
delineated by
HII regions and seen clearly in H$\alpha$ images (Marconi et al. 1994;
Elmouttie et al. 1998a). There is also a molecular disk (Elmouttie et al. 1998c)
or ring (Curran et al. 1998) of radius $\sim$ 300 pc which presumably
provides the gas 
for the star formation. ISO spectra show PAH features between 5 and 
12 $\mu$m which are remarkably similar to those found in `pure' starburst
spectra. There are also lines from H$_{2}$ and low excitation ionic species,
both believed to be associated with
star forming regions (Moorwood et al. 1996). The distribution and kinematics
of the H$_2$\,v=1--0\,S(1) line have been studied from the ground via
imaging (Maiolino et al. 1998; Davies et al. 1998) and long slit
spectroscopy (Storchi-Bergmann et al. 1999).
High resolution Br$\gamma$ 
imaging (Maiolino et al. 1998) has been interpreted in terms of
ongoing star formation
activity within a few tens of pc of the active nucleus.
Maiolino et al. (1998) find that,
between the 100 pc
and 10 pc scales, the stellar population is relatively young with an age 
between 4 $\times$ 10$^{7}$ and 1.5 $\times$ 10$^{8}$ yrs. They
argue that 
the starburst may have propagated outwards from the
nucleus to the 200 pc diameter ring, where hot young stars are currently
forming. The luminosity of the starburst within a 200 pc radius is
$\simeq$ 1.7 $\times$ 10$^{10}$ L$_{\odot}$ (Maiolino et al. 1998),
while the luminosity of the Seyfert nucleus is comparable at $\sim$
10$^{10}$ L$_{\odot}$ (Moorwood et al. 1996). Together, the two contribute
most of the luminosity of the galaxy (Maiolino et al. 1998).

The proximity of
this Seyfert plus circumnuclear starburst galaxy invites further study.
The connection between nuclear activity and circumnuclear star formation is a
topic of great current interest (e.g. Cid Fernandes \& Terlevich 1995;
Genzel et al. 1995; Storchi-Bergmann et al. 1996; Heckman et al. 1997;
Gonz\'alez Delgado et al. 1998) but can only be probed by high spatial 
resolution observations. High spatial resolution imaging is also needed to
investigate the origin and nature 
of the gaseous outflow within the ``ionization cone''.
Such observations can also potentially investigate the morphology of the
gas and dust responsible for the obscuration of the nucleus in type 2 Seyferts.

In this paper, we report an optical and near infrared imaging study of the
Circinus galaxy with the
Hubble Space Telescope,
which provides a resolution better than 0\farcs1 (2 pc) at optical
wavelengths.
We have obtained images in the gaseous emission
lines [OIII] $\lambda$5007, H$\alpha$
and H$_2$\,v=1--0\,S(1) at 2.12 $\mu$m, and
in the green (5470 \AA), red (8140 \AA) and near infrared (2.04 and 2.15
$\mu$m) continua. In addition,
we have obtained an image in the [Fe II] $\lambda$1.644 $\mu$m line with a
ground-based telescope.

\section{Observations and Reduction}

\subsection{HST Optical Observations}

As part of GO program \#7273, the Circinus galaxy was observed by HST
with the WFPC2 instrument on April 10, 1999. This target lies at the
heart of the southern continuous viewing zone (CVZ), allowing us to
obtain 3500 seconds of exposure in a single orbit.  The galaxy was
centered on the PC chip, in order to obtain the maximum spatial
resolution on the nucleus and ionization cone regions.  Images were
obtained through filters isolating the emission lines of
[OIII]$\lambda$5007 (F502N) and H$\alpha$
(F656N), as well as
corresponding green and red continuum regions (filters F547M and
F814W). The exposure times were $2 \times\ 900$ seconds and $2
\times\ 800$ seconds for the \oiii\ and H$\alpha$
imagery, respectively. No
anomalies were reported during the observation period.

The data were reduced in the standard fashion, using the IRAF/STSDAS
software package\footnote{IRAF is distributed by the National Optical
Astronomy Observatories, which are operated by the Association of
Universities for Research in Astronomy, Inc., under cooperative
agreement with the National Science Foundation.}.  Although the data
were received in calibrated form, they were recalibrated and reduced
in order to take advantage of more up-to-date calibration reference
files.  This procedure was accomplished by the CALWP2 task, which
removes the CCD bias level, subtracts the dark current component, and
flat fields the pixel-to-pixel response of the chips.  Cosmic rays
were removed by combining two images in each of the separate filters,
using the anticorrelation technique implemented in the CRREJECT task.
The images were carefully surveyed by hand and additional cosmic rays
identified and removed by interpolation.

The data were flux-calibrated using the SYNPHOT package for synthetic
photometry, together with the most recent HST filter and telescope
throughput tables.  A corresponding continuum image was constructed
for each narrowband image by assuming a linear continuum and
interpolating from the broadband images. The alignment of the images
was then verified using a number of stars in the fields; small offsets
were applied. The continuum was subtracted from each of the narrowband
images to derive maps of the H$\alpha$ and \oiii\ line flux. 
We produced maps encompassing
the entire 160\arcsec\ field of the WFPC2 instrument (0\farcs1
pixel$^{-1}$), as well as the 36\arcsec\ field of the high-resolution
PC chip (0\farcs046 pixel$^{-1}$).

We have also produced a map of the line ratio [OIII]/H$\alpha$ and
of the continuum color V -- I. Each of the [OIII] and H$\alpha$
images was ``clipped'' at a signal to noise (S/N) ratio of 1. The line ratio 
map was calculated at all points where {\it either} map has S/N $>$ 1.
Because H$\alpha$ is seen from all gas which emits [OIII], but
[OIII] is detected from only the high excitation gas, our procedure implies
that the calculated ratio is an actual measurement only where [OIII]
is detected, and an upper limit elsewhere (in HII regions ionized by hot
stars, for example). 

The F656N filter has a peak transmission at 6561\AA\ and a width
$\delta\lambda$ = 21.4\AA\ (as defined in the WFPC2 Instrument Handbook). At
the heliocentric systemic velocity of Circinus
(439 $\pm$ 2 km s$^{-1}$; Freeman et al. 1977), both [NII]$\lambda$6548 and
H$\alpha$ fall on the flat part of maximal transmission in the filter profile. 
Because there is
very little gas with recession velocity below 100 km s$^{-1}$ (see
Fig. 4 of Elmouttie et al. 1998a), the contribution of [NII]$\lambda$6583 to
our F656N image is always negligible. In HII regions
F([NII]$\lambda$6548) $<<$  F(H$\alpha$) and so [NII] is a negligible
contributor to the image in these regions. There will be some contribution
from [NII]$\lambda$6548 in the high excitation gas; this contribution is
difficult to quantify in view of the spatial variation of recession velocity
and the minimal number of published
measurements of the spatial dependence of
F([NII]$\lambda$6548)/F(H$\alpha$). With this caveat, we shall refer to the
continuum-subtracted F656N image as ``H$\alpha$''.

\subsection{HST Infrared Observations}

An additional 4 CVZ orbits were used to observe the Circinus galaxy with the
NICMOS Camera 2 through filters 
F204M, F212N (on-band H$_2$\,v=1--0\,S(1)), and F215N
(off-band H$_2$). A total of 6656\,seconds was spent on the galaxy
in each of the narrow-band filters, using five different dither
positions. These were bracketted by two 128-second on-source
observations in the F204M filter, both taken at the same position. At
the time of designing the NICMOS observations, the temporal stability
of the thermal background was not known. Therefore, three regions of
sky located approximately 10 arcminutes from the center of the galaxy
were also observed to facilitate background subtraction; observations
of 512\,seconds each in the F212N and F215N filters were made at the
start, middle, and end of the visit, and observations of 256\,seconds
with the F204M filter were made at the start and end of the visit.

The first step in the data reduction was to perform a subtraction of
the NICMOS `pedestal' using software written and kindly
provided by R. van der Marel. A study of the background images
revealed no significant variations in either structure or count rate,
and these were therefore combined, using median filtering to remove
Galactic stars, to produce a single background image for each filter.

The images taken at each dither position in the F212N and F215N
filters were stacked and background-subtracted. They were then
combined using the CALNICB task with integer pixel shifts
determined from centroiding of the bright nucleus. The two F204M
images were similarly combined, although no shift was necessary.

As a final step, an image of the H$_2$\,v=1--0\,S(1) emission line
only was constructed by subtracting an estimated continuum level from
the F212N line-plus-continuum image. Several combinations of the F204M
and F215N (which possessed similar signal-to-noise ratios) were
investigated, with there being no significant differences between the
results. We therefore chose to construct a continuum image at the
pivot wavelength of the F212N filter by assuming that the continuum
$S_\lambda$ was
a linear function of $\lambda$ between the F204M and F215N filters.

\subsection{Ground-based Infrared Observations}

Infrared images of Circinus
were obtained through an H-band filter and a 
narrow-band [Fe\,II]$\lambda$1.644$\mu$m filter
using the CIRIM imager at the 4m Blanco telescope at CTIO
on the night of May 25 1999. The total integration times were 180 s for the
H-band 
and 1800 s for the narrow-band observation, which was split into 9 individual exposures. Sky frames were obtained at 10$^{\prime}$
from the galaxy. The scale was 0\farcs21 pix$^{-1}$, 
and the seeing was $\approx$ 0\farcs9.  

Reduction of the ground-based images followed standard procedures in IRAF.
The images were corrected for extinction using average coefficients for Cerro
Tololo (Frogel, 1998).
Because the weather became cloudy before a standard star could be observed,
flux calibration of the H- and narrow-band images was performed using the
aperture photometry of Moorwood \& Glass (1984). 
We have adopted the H-band image as the continuum (off-band)
for the [Fe\,II]$\lambda
1.644\mu$m image, because the bad weather precluded obtaining 
a continuum image through a narrower filter. Two potential problems must be
considered when adopting this procedure. First, the H filter contains the
[Fe\,II]
$\lambda 1.644\mu$m line. From Storchi-Bergmann et al. (1999), we estimate
that the
equivalent width
of the [Fe\,II] $\lambda 1.257\mu$m line (which has comparable
flux to the 1.644$\mu$m line - [FeII]$\lambda1.644\mu$m/1.257$\mu$m =
0.74 - Nussbaumer \& Storey 1988) is $\lesssim$ 5\AA,
much less than the FWHM 
$\approx$3300\AA\ of the H-band filter, and so the contribution of the 
[Fe\,II] emission to the H-band image is negligible. Second, the H filter is 
much wider than the [Fe\,II] filter. However, it is approximately 
centered on the line, so differences in continuum level 
due to color variations should be small.

The continuum-subtracted [Fe\,II] image was obtained by scaling the narrow-band
image with respect to the H-band image. We first tried using the scale 
factor obtained from the ratio between the integrated counts of a field star
in the H image and in the [Fe\,II] image (only one star 
in our images was bright enough for this purpose). The resulting subtracted
image was
negative around the nucleus, indicating that color effects were
present, and that we had oversubtracted the galaxy continuum. We
then progressively increased the scale factor used to multiply the [Fe\,II] 
image 
until the negative counts were eliminated.
This was obtained by a 12\% increase of the
factor obtained from the stellar counts.

\section{Results}

\subsection{Optical Images}

The HST optical images obtained with WFPC2 are shown in Figs 1 - 5. The
orientation is such that the
major axis of the galaxy (p.a. 30$^{\circ}$ $\pm$ 5$^{\circ}$; Freeman et al. 
1977) runs horizontally and the near side (SE) of the galaxy disk 
is in the lower part of the panels.

Fig. 1 shows the large-scale structure of the galaxy in
H$\alpha$ (upper panel) and in the I band
continuum (lower panel). Some 1\farcm5 to the SW of the nucleus
(upper right of figure) in the H$\alpha$ image is a region of
emission which is also apparent in the ground-based H$\alpha$ imaging of
Elmouttie et al. (1998a, their Fig. 1a). Elmouttie et al. note the coincidence
of this feature with compact radio continuum emission and argue that it is
an HII region. The absence of [OIII]$\lambda$5007 at this location in our images
supports this interpretation. Two diffuse patches of line emission, also
apparently HII regions and visible in ground-based images
(Marconi et al. 1994; Elmouttie et al. 1995), are found
$\simeq$ 30$^{\prime\prime}$ S of the nucleus. A linear feature, seen in
ground-based [OIII]$\lambda$5007 (Veilleux \& Bland-Hawthorn 1997) and
H$\alpha$ + [NII] (Marconi et al. 1994) images, extends
$\simeq$ 40$^{\prime\prime}$ to the SW of the nucleus, onto the upper right
WF chip. This feature is not detected in our [OIII]$\lambda$5007 image and is
thus of lower excitation than the high excitation gas (discussed below)
to the NW of the nucleus (cf. Fig. 6 of Marconi et al. 1994). The I band
image reveals a broad, spiral dust band which sweeps S from the intersection of
the PC and upper right WF chips, and then turns E and NE, passing
across the point where the four chips intersect and continuing to the left
hand edge of the lower left WF chip. This dust lane coincides with a region of 
weak H$\alpha$ emission between the two diffuse patches noted above and
the central, bright region of the galaxy. The dust lane may obscure line 
emission in this region (where the four chips meet). 

Fig. 2 shows only the PC chip; the [OIII]$\lambda$5007 image is top left,
H$\alpha$ is top right, V band is lower left and I band is lower right.
Fig. 3 is an color display of the PC image of H$\alpha$.
The [OIII]$\lambda$5007 image
contains a `V' shaped structure (see also Fig. 5) within 2$^{\prime\prime}$
of the nucleus, plus fine scale radial features extending to the NW.
There are also concentrations of bright, high excitation line emission towards
the top of
the panel, with an overall morphology suggestive of an elliptical ring,
which might represent the projection of the end of a tilted, hollow, conical
structure,
with apex near the nucleus. In addition to the emission from HII regions
ionized by hot stars (discussed below), 
the H$\alpha$ image reveals prominent emission from
the [OIII]-emitting gas, including the V-shaped nuclear structure and the
radial features and `blobs' to the NW.
The fainter [OIII]$\lambda$5007 emission
further to the NW, imaged by Veilleux \& Bland-Hawthorn (1997), is
off the top edge of the PC chip\footnote{It was, unfortunately, not possible
to specify a special orientation of HST during these observations in view
of our use of the Continuous Viewing Zone}. 

The H$\alpha$ image (Fig. 2 top right, Fig. 3)
reveals the well-known ring of HII regions of
radius 8 -- 14$^{\prime\prime}$.
The ring shows considerable structure, some of which results from
obscuration. For example, a prominent dust band, best seen in the V band
image (Fig. 2, bottom left), 
extends NE -- SW 2 -- 8$^{\prime\prime}$ S of the nucleus.
The sharp NW edge of this band is seen as a sharp edge in the H$\alpha$
emission some 9$^{\prime\prime}$ SW of the nucleus. Indeed, the
H$\alpha$ line emission
is weaker and more fragmented to the S of the nucleus than in other directions,
apparently a result of obscuration by the dust band. 
Various other dust bands, extending NE - SW to the SE (bottom of panel) of
the nucleus, are seen in the V- and I-band (bottom right) images.
Given the projected shape of this ring of HII regions, 
it cannot be an intrinsically
circular structure in the plane of the galaxy disk.
Close to the nucleus, the H$\alpha$ image also shows an elliptical
ring of HII regions with radius 2$^{\prime\prime}$ towards the NE
(best seen in Figs 5 and 12). We shall refer to this feature as the ``nuclear
ring''. The nuclear ring extends more than half way around
the
nucleus, from S through E to N of the nucleus, with major axis
more or less along the galaxy disk
(NE - SW). This morphology suggests the nuclear ring is intrinsically
circular in
shape, and located in the plane of the galaxy disk (Section 4.2).

The top panel of Fig. 4 is an image of the flux ratio
[OIII]$\lambda$5007/H$\alpha$ over the
PC chip. We should bear in mind that this ratio can only be measured in the
high excitation gas, where
both [OIII] and H$\alpha$ are detected. Elsewhere, in HII regions
where [OIII] is undetected, only an upper limit to the ratio is
obtained (Section 2.1).
This ratio is typically $\lesssim$ 0.5 in the starburst ring, 
confirming the low excitation of the gas and
commensurate with the values expected for HII regions. In the nuclear  
ring of HII regions,
some 2$^{\prime\prime}$ NE of the cone's apex, there is a light band
where the upper limit to the ratio is lower at $\lesssim$ 0.1 - 0.5.
Extending $>$ 4$^{\prime\prime}$ NW from the cone's
apex are seen the high excitation linear streamers and knots which stand out
in the [OIII] image (Fig. 2, top left). 
The ratio for the streamers within 
10$^{\prime\prime}$ of the nucleus is lower than further away; this might be
a result of superposition on background HII regions, 
of reddening close to the nucleus or of a lower ionization parameter due to
a higher gas density near the nucleus. Beyond 10$^{\prime\prime}$,
where the high excitation gas projects outside the starburst ring, the
[OIII]/H$\alpha$ ratio of the streamers and knots ranges from
1 to $\ge$3,
confirming that this gas has Seyfert-type excitation and
suggesting that it is little reddened by dust internal to the Circinus
galaxy. These [OIII]/H$\alpha$ ratios should be multiplied by 1.6 to
correct for the foreground reddening of A$_{V}$ = 1.5 mag 
(Freeman et al. 1977).

The lower panel of Fig. 4 shows the ratio of the F547M and F814W images,
which has been converted to a V -- I color map. This image shows considerable
structure, especially on the near (SE) side of the galaxy disk where red
spiral bands are apparent. Regions as red as V -- I $\simeq$ 3.2 are found.

Enlarged views of the nuclear region (central 
4\farcs0 $\times$ 4\farcs0) are shown in Fig. 5.
In both [OIII] (top left) and H$\alpha$ (top right), a filled V-shaped feature,
with opening angle $\simeq$ 90${^\circ}$, extends
some 2$^{\prime\prime}$ to the NW of the
nucleus (taken as the apex of the `V'). The axis (i.e. the bisector
of the edges) of the `V' is in p.a. $\simeq$ --44${^\circ}$, which is
74${^\circ}$ from the major axis of the galaxy disk and
$\simeq$ 64${^\circ}$ from the optical continuum polarization direction
at the nucleus (Oliva et al. 1998). The fact that the axis of the `V'
is not perpendicular to the polarization is consistent with the conclusion
that the polarization is dominated by transmission
through aligned grains in the galaxy disk, rather than scattering within an
ionization cone (Oliva et al. 1998).
Considerable
structure is found within the `V', including knots, a straight streamer
2$^{\prime\prime}$ W of the nucleus and a curved streamer which extends
first N from the nucleus and then turns W, ending 2$^{\prime\prime}$ from
the nucleus (see also Fig. 12).
The brightest emission within the `V' lies along its southern edge, adjacent
to the detected H$_{2}$ v = 1-0 S(1) emission (Section 3.2.3), perhaps
indicating that the gas density is higher in this region than in other
parts of the `V'.
The bright, compact
feature right at the apex of the `V' is elongated by
$\simeq$ 0\farcs3 in a NNW (p.a. --23${^\circ}$) 
direction. All structures within
the `V' are clearly visible in both lines. The H$\alpha$ image
shows, in addition, emission outside of the `V' which is not seen in [OIII]
and presumably comes from HII regions. 
In particular, the nuclear ring of HII regions with radius
2$^{\prime\prime}$ (discussed above) is apparent, though better seen
in Fig. 12.
The [OIII]/H$\alpha$ ratio
map (Fig. 5, bottom left) confirms that the gas within the `V' has a higher
excitation than that outside of it. Both the straight streamer and the
curved streamer noted above show up as high excitation features. 
The low excitation of the gas in the nuclear ring, some
$\sim$ 2$^{\prime\prime}$ NE of the nucleus,
was noted earlier. The bright, elongated emission-line feature right at the
apex of the `V'
is also seen strongly in the I band continuum
(Fig. 5, bottom right), in which image it
extends 0\farcs7 NNW of the nucleus, at the same p.a.
(--23${^\circ}$) as found for H$\alpha$.

\subsection{Infrared Images}
\subsubsection{The Extended Infrared Continuum and Colors}

The HST continuum image obtained through the F215N filter is shown in
Fig. 6. Diffraction rings are clearly visible in all the HST infrared
images, indicating a strong contribution from an unresolved source
at the nucleus (discussed in section 3.2.2) and confirming the finding of
Maiolino et al. (1998). Fig. 6 also shows the bright circumnuclear
emission plus a series of alternating bright and dark spiral
arms to the SE, which are qualitatively similar to those
seen in the I band image (Figs 1 and 2)

Fig. 7 shows a color image made from our F547M (shown in blue), F814W
(green) and F204M (red) images
and reveals a dramatic color gradient across the galaxy. The strong dust
bands to the SE have already been noted and the
red colors on this side are also indicative of obscuration. A 
color gradient from the near to the far side of the disk
has also been found in H -- K color by Maiolino et al. (1998).

Fig. 8 is a greyscale image of the observed V -- K colors, derived from the
F547M and F204M images. The reddest region is at the nucleus, where 
V -- K $\simeq$ 8.0. The near (SE) side of the galaxy disk is much redder
(shown as darker shades)
than the far (NW) side, as already noted from Fig. 7. The redder
regions on the near side
of the galaxy disk correspond to 6.0 $<$ V -- K $<$ 7.5, while the bluer
extended regions on this side range down to V -- K $\simeq$ 4.2.
On the far (NW) side,
the colors are mostly in the range 4.0 $<$ V -- K $<$ 5.0, although redder
regions (5.0 $<$ V -- K $<$ 6.0) are also present. The foreground extinction 
to Circinus
is A$_{V}$ = 1.5 mag (Freeman et al. 1977),
which corresponds to E(V -- K) = 1.3 mag
(Cardelli, Clayton \& Mathis 1989). Thus the bluest regions have intrinsic
colors V -- K $\le$ 2.7 mag, the equality applying if these regions suffer no
extinction internal to Circinus. Assuming an intrinsic stellar color
of V -- K = 2.7 mag and that the dust lanes on the near side
lie fully in front of these stars, the obscuration through the dust
lanes is typically 2.2 $<$ A$_{V}$ $<$ 3.9 mag.

An intrinsic V -- K = 2.7 mag is bluer than found in bulge
dominated galaxies: the average color for elliptical and lenticular 
galaxies is V -- K $\simeq$ 3.3 (Frogel et al. 1978). The HII regions
and J -- K colors (Storchi-Bergmann et al. 1999) also indicate the presence
of a young stellar population.
The color V -- K = 2.7 mag
is, however, redder than all the starburst models of Leitherer \& Heckman
(1995), except high metallicity, instantaneous bursts with age $\simeq$
10$^{6.9 \pm 0.1}$ yrs. 
It is possible that the foreground extinction of
A$_{V}$ = 1.5 mag underestimates the extinction to the bluest regions
and that their intrinsic V -- K $<$ 2.7.
Alternatively, the observed starlight
could represent a mixture of young disk and old bulge stars.

Any interpretation of the colors must also account
for the strong tendency for dust lanes
to be seen on the near side of the disk of Circinus, and much less so on the
far side.
If all the starlight is from the disk, strong dust bands must lie
between Earth and the stars in the 
galaxy disk on the near side, but not those on the far side.
Outside corotation, the compression takes place on the outer edge of a spiral
arm (since $\Omega_{pattern}$ $>$ $\Omega_{gas}$). On the near (SE) side of
the galaxy disk, the dust lanes associated with a spiral arm would then be
nearer to Earth than the hot stars formed in the arm, an effect with the
correct sense to account for the excess reddening on the near side.
On the other hand, the 
highlighted dust lanes
on the near sides of galaxy disks are usually understood as an effect of the
dusty disk on background bulge light (Hubble 1943). 
This interpretation would require a significant contribution from bulge
light in the inner few hundred pc of Circinus. It is not clear to us which
of these two explanations for the excess of dust lanes and reddening on the
near side of Circinus is the correct one.

\subsubsection{The Unresolved Nuclear Source}

We have used TINYTIM V4.4 to
model the point spread function (PSF) of NICMOS
camera 2. The flux density of the compact nuclear source was then obtained
by two methods. In the first, we used
the IRAF task {\sc sclean} to perform a sigma-CLEAN
deconvolution (Keel 1991). The {\sc tinytim} PSF was computed onto an
oversampled grid of $11 \times 11$ elements per NIC2 pixel and then
resampled to the actual pixel scale so as to allow us to simulate the
source peak being located at different locations within the central
pixel. The sigma-CLEAN algorithm was run for all 121 such resampled
PSFs, and the PSF which gave the largest maximum pixel value in the
final image was assumed to be the most accurate representation of the
true PSF.

The second method involved fitting a de~Vaucouleurs $r^{1/4}$ 
law to the azimuthally-averaged radial surface brightness profile
over the range of radii
$0\farcs6 < r < 5\arcsec$, and then extrapolating this to smaller
radii. We then
scaled the PSF so that the sum of the two components matched the peak
of the surface brightness. These results are shown in
Fig. 9. 

The agreement between the flux densities of the compact nuclear
source derived by the two methods is very good. The
discrepancy exceeds 4\% only in the F212N filter, where
the presence of H$_2$ emission concentrated near the nucleus has
presumably caused the profile fitting method to overestimate the true
contribution of the point continuum source. We opt,
therefore, to use the fluxes
derived from the sigma-CLEAN deconvolution to investigate the nature
of the nucleus, and assume internal errors of 5\% on both the F204M and
F215N measurements. 

However, comparison of our flux-calibrated NICMOS images with
ground-based measurements (Moorwood \& Glass 1984; Glass \& Moorwood
1985) indicates a systematic discrepancy between the two sets of
photometry. Our F212N and F215N images are brighter by 31\% and 32\%,
respectively, in all apertures, while the nucleus contributes only
4--10\% of the total light, depending on the aperture size. Our F204M
image is brighter by 23\% assuming a flat spectrum in the $K$-band,
although the true discrepancy will be slightly larger as the flux
density of Circinus increases with wavelength (Storchi-Bergmann et al.\
1999). There thus appears to be a systematic offset of $\sim 30$\%
in the NIC2 photometry for these three filters. We have, therefore, reduced
all flux measurements by 30\% and the resulting flux densities of the
point source are listed in Table 1. Our estimate of the point source flux is
in line with that of Maiolino et al. (1998), who found 24 mJy at K
($\lambda_{eff}$ = 2.2$\mu$m).

The observed ${\rm F215N} - {\rm F204M}$ color can be reproduced by a power law
with a spectral index $\alpha_{o} = 7.0 \pm 1.2$
(S$_{\nu} \propto \nu^{-\alpha}$); the L$^{'}$ (3.8 $\mu$m)
flux measured by Maiolino et al. (1998) agrees with the
extrapolation of this spectrum. If we assume that this
steep spectrum is the result of a heavily-reddened, much shallower power
law, it is possible to infer the obscuration, A$_{V}$, as a function of
the intrinsic spectral index, $\alpha_{i}$, as shown in Fig. 10. Fadda et al.
(1998) find that the spectral index of Seyfert 1 galaxies at K band
is $\alpha_{i} \simeq 1.8$. If the unresolved infrared source in the nucleus of
Circinus is actually an obscured Seyfert 1 nucleus with this value of
$\alpha_{i}$, then the foreground extinction is
$A_V = 28 \pm 7$ mag.
This value is in excellent
agreement with the optical depth of the 9.7\,\micron\ silicate feature
observed by Moorwood \& Glass (1984), assuming an underlying blackbody
spectrum and $A_V/\tau_{9.7} = 16.6$ (Rieke \& Lebofsky 1985), and is
also consistent with the $A_V \sim 20$\,mag inferred by Marconi et
al.\ (1994). Applying a correction of
$A_V = 28$ mag, the unobscured 2.2\,\micron\
flux of the nucleus would be 446\,mJy.

The extinction to the near-infrared nucleus is much lower than
expected from the gas column of 4 $\times$ 10$^{24}$ cm$^{-2}$ to the nuclear
X-ray source
(Matt et al. 1999). Assuming a normal gas-to-dust ratio, the implied extinction
is $A_V \sim 2,000$ mag or $A_K \sim 230$ mag. This difference 
indicates that the near-IR 
continuum originates in regions farther from the nucleus
than the X-rays, most probably being re-emission of nuclear radiation by dust.
As the near-IR source
is unresolved, the radius of the re-emitting structure is smaller than
2 pc and could be the inner regions of a torus. Alternatively, the inner
regions of the torus may have an obscuration similar to that of the
X-ray source and thus be invisible in the near infrared. The observed
K band light could then originate from hot dust in the inner part of the
narrow line region,
perhaps along the polar axis of the torus and heated by nuclear optical-uv
light or mass outflow, as may be the case in NGC 1068
(Braatz et al. 1993;
Weinberger et al. 1999). Such a geometry provides a natural explanation for the
lower obscuration to the near infrared source than to the hard X-ray source.
Other explanations for a large N$_{H}$/$A_{V}$
ratio found in a Seyfert 2 galaxy are discussed in Simpson (1998).

The absorption-corrected K-band and 2-10 keV luminosities of the compact
nucleus are L(K) $\simeq$ 2.4 $\times$ 10$^{41}$ erg s$^{-1}$ and
L(2-10 keV) = 3.4 $\times$ 10$^{41}$ -- 1.7 $\times$ 10$^{42}$ erg s$^{-1}$
(Matt et al. 1999), respectively. These numbers agree well with an 
extrapolation to lower luminosities of the correlation between L(K) and
L(2-10 keV) for hard X-ray selected active galaxies, mainly Seyfert 1s
(Kotilainen et al. 1992, their Fig. 7c). This agreement supports the view
that Circinus contains a Seyfert 1 nucleus with different obscuring columns
to the K band and hard X-ray sources. It also indicates that the hot dust
radiating in the near infrared ``sees'' the dust-heating source (i.e. the
compact optical, uv and X-ray source) through a smaller column of gas than
we do.

\subsubsection{The Molecular Hydrogen Emission}

The continuum-subtracted image of the H$_2$\,v=1--0\,S(1) 
emission line is shown in Fig. 11.
Unfortunately, the strong contribution from the point
source results in significant residual artifacts at and near the
nucleus. Attempts to remove these, by subtracting a scaled PSF from
the images before combining them, did not prove successful, and we are
therefore unable to learn anything about the structure of the H$_2$
emission closer than 0\farcs5 from the nucleus.

The molecular hydrogen extends towards the WSW of the nucleus, in agreement
with the inner part of the image in the same line presented by
Maiolino et al. (1998). We fail to detect the much fainter, more
extended emission seen by Maiolino et al. (1998), Davies et al. (1998)
and Storchi-Bergmann et al. (1999). The H$_{2}$ is strongest
outside, and to the S, of the V-shaped region of high excitation ionized
gas (compare Figs 5 and 11, which have the same field size and
orientation),
confirming the recent findings of Maiolino et al. (2000) from a ground-based
image in the same line.

The molecular hydrogen
emission $\simeq$ 0\farcs75 to the WSW of the nucleus is 
at least
three times brighter than that at the same distance ENE of
the nucleus. If the
distribution is intrinsically symmetric, but the eastern emission is
obscured, a minimum of 1.2\,magnitudes of extinction at 2.12\,\micron\
would be required, corresponding to $A_V \gtrsim 10$\,mag. This is
consistent with the extinction we derived to the nuclear infrared continuum
source (Section 3.2.2), and so we
may be observing the effects of a large-scale obscuring structure
close to the plane of the galactic disk (McLeod \& Rieke 1995; Simcoe
et al.\ 1997). Alternatively, of
course, there may simply be more molecular gas, or its temperature and/or
density may be more conducive to emission in the H$_2$\,v=1--0\,S(1) line,
to the west of the nucleus than to the east.

\subsubsection{The [Fe II] Emission}

A contour map of the [Fe II]$\lambda$1.644 $\mu$m line is shown superposed
on the HST H$\alpha$ image in Fig. 12. There is a compact source
(flux 3.2 $\times$ 10$^{-14}$ erg cm$^{-2}$ s$^{-1}$) , which
we presume coincides with the nucleus, plus an extended arc-like feature
(flux 1.8 $\times$ 10$^{-14}$ erg cm$^{-2}$ s$^{-1}$) to
the NE. Faint [Fe II] emission is also associated with a spiral arm some
5$^{\prime\prime}$ - 7$^{\prime\prime}$ to the E and SE of the nucleus.
Our image and fluxes are broadly 
consistent with those obtained by Davies et al. (1998) in the
same line, though their resolution was 
lower and the arc-like feature is seen in
their image as a small extension
of the nucleus to the NE. This arc-like feature
coincides with, and is morphologically similar to, the NE part of the
nuclear ring
of HII regions seen in H$\alpha$ (Section 3.1). 
The [Fe II] emission in the arc then 
presumably originates from shock excitation in
supernova remnants. A weak radio continuum source coincides with the [Fe II]
arc (Davies et al. 1998). 
The [Fe II] luminosities of the arc and the nucleus lie somewhat below the
extrapolation to lower luminosity of the best fit to the empirical
correlation between
[Fe II] and radio luminosity found by Forbes \& Ward (1993) and
Simpson et al. (1996) for Seyferts
and starbursts. However, the Circinus values are consistent with the
correlation given its scatter.
The observed 3$^{\prime\prime}$ (60 pc) length
of the arc rules out the suggestion (Davies et al. 1998) 
that all the [Fe II] emission comes from a single, young supernova remnant.
However, Maiolino et al. (2000) find that the brightest [Fe II] emission of
the arc is compact in HST images, with a size $<$ 4 pc, and argue that it may
trace a single supernova remnant.

The origin of the [Fe II] associated with the nucleus itself is less clear,
but Storchi-Bergmann et al. (1999) have found that the
[Fe II]$\lambda$1.257/Pa$\beta$ ratio is $\simeq$ 0.4 at the nucleus and
increases outwards. This low nuclear [Fe II]$\lambda$1.257/Pa$\beta$ ratio
is typical of starbursts (Colina 1993; Simpson et al. 1996), so
it may be that all the
[Fe II] emission in the inner regions of Circinus 
- nucleus, arc and spiral arm -
is dominated by supernova remnants. This result may be contrasted with
the situation in the nucleus of NGC 1068, where Blietz et al. (1994)
have found that the [Fe II] emission is elongated along the radio
jet, and may be powered by jet-driven shocks or photoionized by the active 
nucleus.

\section{Discussion - Gas in The Inner 100 pc of the Circinus Galaxy}

\subsection{The V-shaped High Excitation Gas}

We have discovered a compact ($\simeq$ 30 pc), sharply-bounded, filled V-shaped
region of high excitation gas (Figs 5 and 12). This gas extends to the NW and 
projects in the same direction as the larger scale ($\sim$ 500 pc)
`ionization cone' (Marconi et al. 1994; Veilleux \& Bland-Hawthorn 1997;
Elmouttie et al. 1998a). Both the large scale and small scale `cones'
undoubtedly extend out of the galaxy disk and are visible because
they project against the far side of the disk; any counter-cone is obscured
by dust in the disk. The opening angle of the compact `V' is $\simeq$
90${^\circ}$ and the projection of its edges to larger radii fully
envelop the more extended high excitation gas. It is thus entirely possible
that the compact `V' represents the projection of a `classical
ionization cone' - ionizing photons escaping anisotropically from an
unresolved
source of ionizing photons. This collimation can result from either
shadowing of an isotropic ionizing source by an optically thick torus of
gas and dust or intrinsic anisotropy of the continuum source
(e.g. Wilson 1992). The V-shaped structure extends down to our
resolution ($\le$ 0\farcs1, see Figs 5, 12), so the collimation must occur on
scales $<$ 2 pc. Krolik \& Begelman (1988) state that
the position of the inner edge of the torus is determined by a balance
between the inward flow of clouds and the rate at which the nuclear
continuum can evaporate them. They estimate that the radius of the inner
edge is a few times $S$, where
\begin{equation}
S=0.26L_{44} {{L} \overwithdelims () {L_{E}}}^{-{3}}
{{T_{C} \overwithdelims () {3 \times 10^{7} K}}}^{-2} \rm{pc}
\end{equation}
\noindent
Here $L_{44}$ is the central luminosity in units of 10$^{44}$ erg s$^{-1}$,
$L_{E}$ is the Eddington luminosity and $T_{C}$ is the Compton
temperature. According to Moorwood et al. (1996), the total
luminosity of the Seyfert nucleus is $\simeq$ 4 $\times$ 10$^{43}$
erg s$^{-1}$.
If $T_{C}$ = 3 $\times$ 10$^{7}$ K, then
$L/L_{E}$ must be $\gtrsim$ 0.7 given our upper limit of 1 pc for the inner
radius, which
we have taken to be 3 $\times$ $S$.
This means that the mass of the central black hole would have to be
M$_{BH}$ $\le$ 5 $\times$ 10$^{5}$ M$_{\odot}$ for Krolik \& Begelman's
description to be valid, a rather stringent limit. If 
$T_{C}$ $\le$ 3 $\times$ 10$^{7}$ K, the upper limit becomes lower.
The limit on M$_{BH}$ is stricter than the dynamically
measured upper limit to the
black hole mass of
M$_{BH}$ $\le$ 4 $\times$ 10$^{6}$ M$_{\odot}$ (Maiolino et al. 1998).

There are clear indications that the simplest picture - an anisotropic
ionizing source illuminating ambient gas - is inadequate. First,
kinematic measurements (Veilleux \& Bland-Hawthorn 1997;
Elmouttie et al. 1998a) show that the gas in the large-scale cone is
outflowing. Radio maps show polar radio lobes fueled by outflow from
the nuclear region (Elmouttie et al. 1998b). Second, our [OIII]
image suggests that the high excitation gas some 200 - 400 pc from the nucleus
takes the form of a roughly elliptical annulus (Fig. 2, upper left),
suggesting we are viewing the open end of an inclined, circular conical
structure extending roughly perpendicular to the galaxy disk. The cone at this 
radius seems to be hollow in the [OIII]
image, presumably as a result of a low density outflowing wind or radio jet
which has entrained dense gas along its edges, perhaps from the same
compact dense torus which is supposed to collimate the ionizing radiation.
Third, the existence of this outflow is a potential source of ionizing
photons through photoionizing shocks (e.g. Dopita \& Sutherland 1995).
The outflow velocities observed - 150 -- 200 km s$^{-1}$ - are, however,
rather low for production of a significant ionizing luminosity, but
higher velocity, lower density gas may also be present.

It is in principle possible that `V' shaped structures, like that
seen in Circinus, simply represent gas seen along low-obscuration lines of
sight, with high obscuration to either side.
We do not favor this idea because
of the alignment of the compact `V' with the larger scale,
high excitation gas, which is essentially unobscured. The
region between the obscuring structures (i.e. the compact `V')
would have to be aligned by chance with the larger scale gas in such
a picture. We thus conclude
that the compact `V' is either a classical (ionization-bounded)
ionization cone or a matter-bounded outflow with impressively straight edges.

\subsection{The Compact Circumnuclear Star-Forming Ring}

Our H$\alpha$ image shows the well known ring of HII regions
with radius 150 - 270 pc (Figs 2 and 3). This image also reveals a much more
compact (radius 40 pc towards the NE), clumpy structure of HII regions,
which appear to form more than half of a nuclear ring
(Section 3.1, Figs 5 and 12). All of the compact `V' of high excitation gas
projects
inside the ring. 
However, the ring is not exactly centered on the apex of the 
`V'.
The major axis of the nuclear ring lies in
p.a. $\sim$ 30$^{\circ}$, close to the major axis of the disk. Further,
the major-to-minor axial ratio of the approximately elliptical
ring is $\sim$ 2.3, which 
corresponds to an inclination of $i_{r} \sim 64^{\circ}$ if the ring is
intrinsically circular. This inclination is identical to the
inclination of the large scale stellar
disk - $i_{d} = 65^{\circ} \pm 2^{\circ}$ (Freeman et al. 1977). 
This agreement between both the major axis
and inclination for the compact 
ring and large scale stellar disk indicates that we are
dealing with an intrinsically circular ring in the plane of the disk.
The two star-forming rings probably represent dynamical
resonances (e.g. Buta 1995) and may result from gas being forced from further
out in the galaxy towards the nucleus by oval distortions or bars, such
as the 100 pc-long gaseous bar claimed by Maiolino et al. (2000).

It is also of interest to ask how close to the nucleus we see HII regions.
The southern part of the ring projects to within 5 pc of the apex of the 
ionization cone (Fig. 12).
However, if our interpretation is correct, the inner edge of the nuclear ring
is actually at least 15 pc from the apex. There is fainter
H$\alpha$ emission between the ring and the cone's apex, but
it is rather smoothly distributed
and could be gas ionized by the hot stars which are presumably present in
the ring. We find no conclusive evidence for hot stars within $\sim$ 10 pc
of the nucleus, so models of starbursts within this region (Maiolino
et al. 1998) should be treated with caution.

\section{Conclusions}

Our optical and infrared HST images of the Circinus galaxy reveal a wealth of
detail. There is a sharp morphological distinction between the high 
excitation ionized gas associated with the Seyfert activity and the low
excitation gas in HII regions ionized by hot stars. The infrared image in
H$_{2}$ v = 1-0 S(1) reveals the morphology of the warm molecular gas 10 - 20
pc from the nucleus. Our continuum images at V, I and K bands provide
information on the stellar distribution, obscuring dust features and the
compact nuclear continuum source. We discuss each of these components in turn.

\noindent
{\it 1. The high excitation gas.} On the smallest resolvable scales, there is
a very bright feature extended by 0\farcs3 (6 pc, see Fig. 5). It is at the
apex of a compact (2$^{\prime\prime}$ $\simeq$ 40 pc), filled,
V-shaped region of emission, which is presumably conical or wedge-shaped in
3 dimensions. The coincidence of the bright feature with the cone's apex
suggests that it marks the position of the nucleus. The filled V-shaped 
structure has an opening angle of $\simeq$ 90$^{\circ}$ and is brightest on its
S side, adjacent to the brightest H$_{2}$ v = 1-0 S(1) emission. 
The filled `V' may 
be a `classical' ionization cone (i.e. an ionization-bounded structure 
illuminated by an anisotropic nuclear source) or a matter-bounded region with
impressively straight edges. The length of the `V' ($\sim$ 40 pc) is
consistent with expectations for the half thickness of the galaxy's gas
disk, so we are observing the interaction of the Seyfert nucleus with the
normal gas disk, before the ionizing radiation and outflow break out into
the galactic halo.
Extension of the arms of the `V' to larger radii
encompasses the high excitation gas previously imaged in ground-based
observations. This alignment is consistent with the ionization-bounded
interpretation in which anisotropic nuclear radiation photoionizes the high
excitation gas on all scales. The structure which collimates the ionizing
photons is unresolved by our observations and must thus be $<$ 2 pc in
extent. At 10$^{\prime\prime}$ - 20$^{\prime\prime}$ (190 - 380 pc) from the
nucleus, the morphology of the high excitation gas is suggestive of an
elliptical ring (Fig. 2), which we interpret as emission from the edges of
a circular conical structure viewed at an oblique angle. This concentration
of gas on the edge of the cone suggests entrainment of dense gas, presumably 
in the disk of the galaxy, which is then carried out of the galactic disk by
a lower density wind, which may also power the radio lobes. 

\noindent
{\it 2. The HII regions.} On the smallest scales, we find a partial
elliptical ring (Figs 5 and 12), dubbed the nuclear ring, of HII regions of
radius 2$^{\prime\prime}$ (40 pc) surrounding the projection of the compact
high excitation cone. The major axis position angle and axial ratio of
the nuclear ring are the same as those of the galaxy disk, so the ring is an 
intrinsically circular structure in the disk plane. Some of the [Fe II]
$\lambda$1.644$\mu$m emission coincides with this ring and is thus
presumably emitted by supernova remnants. The nature of the more compact
[Fe II] emission, close to the nucleus, is less clear.

Our images also show the well known ring of HII regions of radius
8$^{\prime\prime}$ - 14$^{\prime\prime}$ (150 - 270 pc, Figs 2, 3 and 4)
in unprecedented detail. The apparent major axis of this ring is inclined by
$\sim$ 30$^{\circ}$ from the disk major axis and it is rounder than the
disk isophotes. If coplanar with the galaxy disk, it must be elliptical in 
shape. Dust bands strongly influence the morphology of this ring,
particularly to the S of the nucleus.

\noindent
{\it 3. The warm molecular hydrogen.} The H$_{2}$ v = 1-0 S(1) image
shows emission extending some 1$^{\prime\prime}$ (20 pc) from the nucleus
(Fig. 11). This emission abuts the S edge of the compact cone of ionized gas,
suggesting a continuous gaseous structure of which part is ionized and
part
molecular. The molecular gas is apparently warmed by the nuclear source 
to the temperatures needed for H$_{2}$ v = 1-0 S(1) emission.

\noindent
{\it 4. Disk starlight.} As expected, dust bands and reddening are most
prominent on the near (SE) side of the galaxy disk (Figs 1, 2, 4, 6, 7 and
8). The bluest  
(in V -- K) regions of the disk are bluer than found in bulge-dominated
galaxies, apparently because of the presence of young stars.

\noindent
{\it 5) The compact nuclear infrared source.} Our observations confirm the
existence of a compact ($<$ 2 pc), very red ($\alpha$ $\simeq$ 7.0,
S $\propto$ $\nu^{-\alpha}$) continuum source in the K band (Figs 6 and 8).
It is proposed that this source is a Seyfert 1 nucleus obscured by
A$_{V} = 28 \pm 7$ mag of extinction. This extinction is smaller than expected
from the X-ray absorbing column (N$_{H}$ = 4 $\times$ 10$^{24}$ cm$^{-2}$)
and a normal gas-to-dust ratio, and we have argued that the infrared source is
indeed
observed through a smaller column of gas than the X-ray source.

This research was supported by STScI and NASA through grants GO7273 and
NAG 81027, respectively. 
We wish to thank the staff of the Space Telescope Science
Institute for their hospitality and especially Christine Ritchie and
Roeland van der Marel for useful discussions. We thank Sylvain Veilleux for
valuable comments on the manuscript.

\vfil\eject

\clearpage

\figcaption[fig1.ps]
{HST images of the Circinus galaxy, showing the whole field
covered by the PC and WF camera chips. The image received on the PC
chip has been binned to the WF pixel size of 0\farcs0996.
Long tick marks are separated by
1$^{\prime}$ and short tick marks by 10$^{\prime\prime}$. A long tick
mark on each axis is aligned with the apex of the ``ionization cone''.
The whole field of each panel is 148$^{\prime\prime}$ $\times$
148$^{\prime\prime}$. The direction of north is rotated +61.8$^{\circ}$
from the vertical and is indicated by the arrow. Darker shades represent
brighter regions. The shading in both images is on a logarithmic scale.
{\it Top panel}. Continuum-subtracted
H$\alpha$ image. The shading 
ranges between 3  $\times$ 10$^{-18}$ (white) and 5 $\times$ 10$^{-16}$ (black)
erg cm$^{-2}$ s$^{-1}$ pixel$^{-1}$. {\it Bottom panel}. Image through
F814W filter (continuum).
The shading ranges between
1 $\times$ 10$^{-16}$ (white) and 2 $\times$ 10$^{-14}$ (black)
erg cm$^{-2}$ s$^{-1}$ pixel$^{-1}$ in the filter band.
\label{Figure 1}}

\figcaption[fig2.ps]
{HST images of the Circinus galaxy, showing the PC chip only.
The pixel size is 0\farcs0455.
Long tick marks are separated by 10$^{\prime\prime}$ and short
tick marks by 2$^{\prime\prime}$.
A long tick
mark on each axis is aligned with the apex of the ``ionization cone''.
The field of each panel is
34\farcs0 $\times$
34\farcs0.
The direction of north is rotated +62.2$^{\circ}$
from the vertical and is indicated by the arrow. Darker shades represent
brighter regions. The shading in all images is on a logarithmic scale. 
{\it Top left panel.} Continuum-subtracted [OIII]$\lambda$5007 image.
The shading ranges between 
8 $\times$ 10$^{-18}$ (white) and 3 $\times$ 10$^{-16}$ (black)
erg cm$^{-2}$ s$^{-1}$ pixel$^{-1}$.
{\it Top right panel.} Continuum-subtracted
H$\alpha$ image. 
The shading ranges between 
2 $\times$ 10$^{-18}$ (white) and 3 $\times$ 10$^{-16}$ (black) 
erg cm$^{-2}$ s$^{-1}$ pixel$^{-1}$. 
{\it Lower left panel.} Image through F547M filter (continuum). 
The shading ranges between
6 $\times$ 10$^{-17}$ (white) and 4 $\times$ 10$^{-15}$ (black)
erg cm$^{-2}$ s$^{-1}$ pixel$^{-1}$ in the filter band.
{\it Lower right panel.} Image through F814W filter (continuum).
The shading ranges between  
1.5 $\times$ 10$^{-16}$ (white) and 1.5 $\times$ 10$^{-14}$ (black)  
erg cm$^{-2}$ s$^{-1}$ pixel$^{-1}$ in the filter band.
\label{Figure 2}}

\figcaption[fig3.ps]
{HST color image of the H$\alpha$ emission of the Circinus galaxy, showing the
PC only. Long tick marks are separated by 10$^{\prime\prime}$ and short
tick marks by 2$^{\prime\prime}$.
A long tick
mark on each axis is aligned with the apex of the ``ionization cone'', which 
can be seen as the brightest region in the center of the image. The field and
orientation are the same as in Fig. 2.
\label{Figure 3}}

\figcaption[fig4.ps]
{HST images of the Circinus galaxy, showing the PC chip only (see Fig. 2
for details of pixel size, scale and orientation). 
{\it Top panel.} The [OIII]$\lambda$5007 image
divided by
the H$\alpha$ image.
Darker shades represent higher ratios.
The shades cover the range of ratios --1 (white) to 2 (black)
on a linear scale.
{\it Bottom panel.} V -- I color image. The  shades range between
V -- I = 1.7 (white) and 3.2 (black) and are linear in magnitude.
Darker shades thus represent redder colors. Locations where the (noisier)
V band image has negative flux are colored white.
\label{Figure 4}}

\figcaption[fig5.ps]
{HST images with the PC of the nuclear region of the
Circinus galaxy. Long tick marks are separated by         
1$^{\prime\prime}$ and short tick marks by 0\farcs2.
A long tick mark on each axis is aligned with the apex of the ``ionization
cone''. The field of each panel is 4\farcs0 $\times$
4\farcs0. The orientation and pixel size are 
the same as in Fig. 2. Darker
shades represent higher values in all panels and all shading is on a linear
scale.
{\it Top left panel.} Continuum-subtracted [OIII]$\lambda$5007 image.
The shading ranges between
1 $\times$ 10$^{-18}$ (white) and 3 $\times$ 10$^{-16}$ (black)
erg cm$^{-2}$ s$^{-1}$ pixel$^{-1}$.
{\it Top right panel.} Continuum-subtracted
H$\alpha$ image.
The shading ranges between
1 $\times$ 10$^{-18}$ (white) and 3 $\times$ 10$^{-16}$ (black)
erg cm$^{-2}$ s$^{-1}$ pixel$^{-1}$.
{\it Bottom left panel.} The [OIII]$\lambda$5007 image
divided by
the H$\alpha$ image. The ratio is calculated
for all areas with S/N $\ge$ 1 in both images.
The ratio ranges from --1 (white)
to 1.5 (black).
{\it Bottom right panel.} Image through F814W filter (continuum).
The shading ranges between
5 $\times$ 10$^{-16}$ (white) and 7 $\times$ 10$^{-15}$ (black)
erg cm$^{-2}$ s$^{-1}$ pixel$^{-1}$ in the filter band.
\label{Figure 5}}

\figcaption[fig6.ps]
{HST image of the near infrared continuum of the Circinus galaxy obtained
with NICMOS Camera 2 and filter F215N. The pixel size was rebinned to that
of the PC (0\farcs0455) and
the main image shows the full field (19\farcs2 $\times$ 19\farcs2) of the
camera. Long tick marks are separated by 10$^{\prime\prime}$ and short tick
marks by 2$^{\prime\prime}$. The orientation is the same as Fig. 2.
Darker shades represent brighter regions. The shading in the main panel 
is on a linear scale and ranges between 0 (white) and
1.4 $\times$ 10$^{-14}$ erg cm$^{-2}$ s$^{-1}$ $\mu$m$^{-1}$ (PC pixel)$^{-1}$
(black).
The inset shows the area of the white dotted square with shading on a
logarithmic scale between
0.6 (white) and 5.0 (black) 
$\times$ 10$^{-14}$ erg cm$^{-2}$ s$^{-1}$ $\mu$m$^{-1}$ (PC pixel)$^{-1}$.
The peak flux is
68 $\times$ 10$^{-14}$ erg cm$^{-2}$ s$^{-1}$ $\mu$m$^{-1}$ (PC pixel)$^{-1}$
The tick marks in the inset are separated by 1$^{\prime\prime}$.
\label{Figure 6}}

\figcaption[fig7.ps]
{Optical - infrared color image of Circinus obtained from HST images through
filters F547M (approximate V), F814W (approximate I) and F204M (within the
K band). The image is 19\farcs2 on a side. 
Long tick marks are separated by 10$^{\prime\prime}$ and short tick
marks by 2$^{\prime\prime}$. The orientation is the same as Fig. 2.
The dark circle in the lower right is
the
coronographic hole. The color scheme is such that F547M is mapped to blue,
F814W to green and F204M to red. 
No correction has been applied for foreground 
reddening. 
\label{Figure 7}}

\figcaption[fig8.ps]
{Grey scale image of the V -- K colors of Circinus obtained from the F547M and
F204M images. 
No correction has been applied for foreground 
reddening. Darker regions are redder; the reddest region is at the
nucleus and has V -- K $\simeq$ 8.0, while
the bluest region is to the NW and has V -- K $\simeq$ 4.0. 
The orientation is the same as Fig. 2.
\label{Figure 8}}

\figcaption[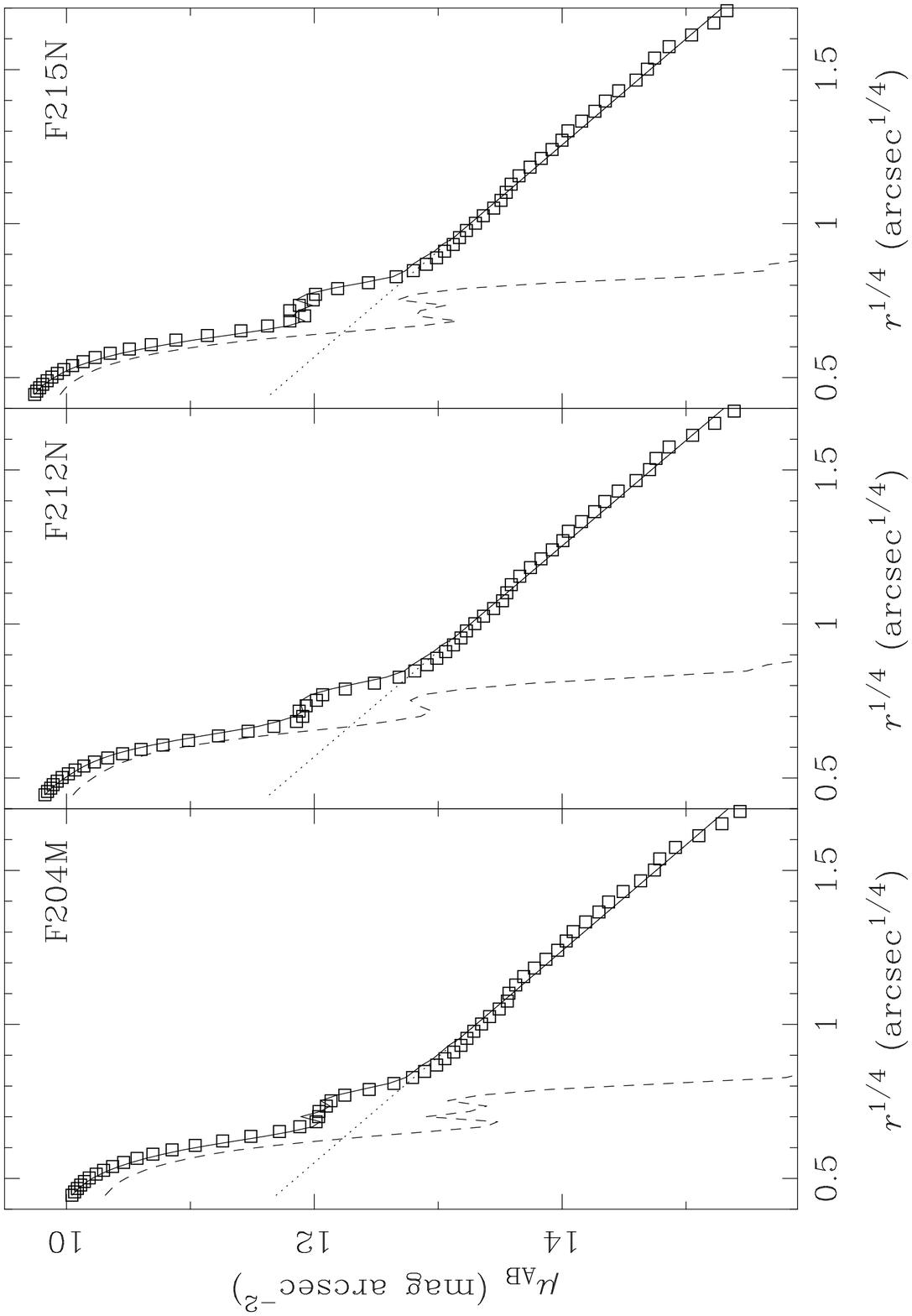]
{Radial surface brightness profiles of Circinus in the three
NICMOS filters. The dotted lines represent the results of fitting a
de~Vaucouleurs $r^{1/4}$ law to the region $0\farcs6 < r < 5\arcsec$,
and the dashed line is the point spread function scaled so that the
sum of the two components (the solid line) matches the peak of the
observed emission. The flux density scale has not been corrected for the
suspected systematic error in the NICMOS fluxes (see text), and thus may
be 30\% too high.
\label{Figure 9}}

\figcaption[fig10.ps]
{Plot of visual obscuration, A$_{V}$ (left axis and solid line), against 
intrinsic spectral 
index, $\alpha_{i}$, for the nucleus of Circinus.
The right axis and dashed line
show the corresponding intrinsic flux density at 2.2$\mu$m.
This flux density scale has not been corrected for the
suspected systematic error in the NICMOS fluxes (see text), and thus may
be 30\% too high. The 
extinction law of Cardelli, Clayton
\& Mathis (1989) has been used.
\label{Figure 10}}

\figcaption[fig11.ps]
{Image of the H$_2$\,v=1--0\,S(1) line, obtained by subtracting the continuum
from the F212N image (see text). 
Long tick marks are separated by 1$^{\prime\prime}$
and short tick marks by 0\farcs2. The field of view and orientation are the
same as in Fig. 5. Darker shades represent brighter regions.
Structure inside the circle, which is centered on the bright,
nuclear continuum source, is unreliable (see text).
\label{Figure 11}}

\figcaption[fig12.ps]
{Contours of [Fe II]$\lambda$1.644 $\mu$m emission superposed on a
grey scale of the HST
H$\alpha$ image. Long tick marks
are separated by 5$^{\prime\prime}$ and short tick marks by 
1$^{\prime\prime}$. The field of view is 13\farcs7 $\times$ 13\farcs7
and the orientation is the same as Fig. 2. Darker shades represent 
brighter regions, and the shading
is on a linear scale from 2 $\times$ 10$^{-18}$ (white) to
8 $\times$ 10$^{-17}$ (black) erg cm$^{-2}$ s$^{-1}$ PC pixel$^{-1}$. The
contours
of [Fe II] emission are plotted at (5, 10, 15, 20, 40, 60, 80 and 100)
$\times$ 1.69 $\times$ 10$^{-16}$ erg cm$^{-2}$ s$^{-1}$ (arc sec)$^{-2}$.
The images
were superposed by aligning their peaks; this alignment is arbitrary.
The arc of [Fe II] emission 2$^{\prime\prime}$ NE of the nucleus is
associated with an elliptical ring of HII regions with observed major axis
along the galaxy disk.
\label{Figure 12}}

\clearpage

\begin{table}
\caption[]{Flux densities of the nuclear component in the three NICMOS
filters, derived using SCLEAN, and by fitting a point source
plus stellar bulge to the observed radial surface brightness profile. A
correction for the apparent systematic error in the flux calibration of
NIC2 has been applied (see text).}
\label{tab:nuke}
\begin{center}
\begin{tabular}{lccc}
\hline
& & \multicolumn{2}{c}{Nuclear flux (mJy)} \\
Filter & $\lambda_{\rm eff}$ (\micron) & {\sc sclean} & Profile \\
\hline
F204M & 2.0355 & 15 & 15 \\
F212N & 2.1213 & 21 & 23 \\
F215N & 2.1487 & 21 & 22 \\
\hline
\end{tabular}
\end{center}
\end{table}

\end{document}